\documentclass{llncsmod01}

\usepackage{latexsym}
\usepackage{amssymb}
\usepackage{graphics}
\usepackage{graphicx}
\usepackage{amsfonts}
\usepackage{amssymb}
\usepackage{url}
\usepackage{subfigure}

\begin{document}

\title{Logic gates and complex dynamics in a hexagonal cellular automaton: the Spiral rule}

\author{Rogelio Basurto\inst{1,2}, Paulina A. Le\'on\inst{1,2} \\ Genaro J. Mart{\'i}nez\inst{1,3}, Juan C. Seck-Tuoh-Mora\inst{3}}

\institute{Escuela Superior de C\'omputo, Instituto Polit\'ecnico Nacional, M\'exico. \\
\and
Centro de Investigaci\'on y de Estudios Avanzados del Instituto Polit\'ecnico Nacional, M\'exico. \\
Email: \url{{larckov, pauanana}@gmail.com} \\
\and
Unconventional Computing Group, Computer Science Department, University of the West of England, United Kingdom. \\
Email: \url{genaro.martinez@uwe.ac.uk} \\
\and
\'Area Acad\'emica de Ingenier{\'i}a, Universidad Aut\'onoma del Estado de Hidalgo, Pachuca, Hidalgo, M\'exico. \\
Email: \url{jseck@uaeh.edu.mx}
}


\maketitle

\begin{abstract}
In previous works, hexagonal cellular automata (CA) have been studied as a variation of the famous Game of Life CA, mainly for spiral phenomena simulations; where the most interesting constructions are related to the Belousov-Zhabotinsky reaction. In this paper, we analyse a special kind of hexagonal CA, {\it Spiral rule}. Such automaton shows a non-trivial complex behaviour related to discrete models of reaction-diffusion chemical media, dominated by spiral guns which easily emerge from random initial conditions. The computing capabilities of this automaton are shown by means of logic gates. These are defined by collisions between mobile localizations. Also, an extended classification of complex self-localisation patterns is presented, including some self-organised patterns. \\

{\bf Keywords}: hexagonal cellular automata, Spiral rule, logic gates, localizations, collisions

\end{abstract}

\noindent {\small Published in {\it Journal of Cellular Automata}, vol. 8, num. 1-2, p. 53-71, 2013. URL: \url{http://www.oldcitypublishing.com/journals/jca-home/jca-issue-contents/jca-volume-8-number-1-2-2013/jca-8-1-2-p-53-71/}}

\section{Antecedents}
Spiral rule is a synchronous totalistic three-state, two-dimensional hexagonal CA introduced by Adamatzky and Wuensche in 2005 \cite{kn:WA06}. This automaton has a complex behaviour dominated by mobile and stationary localizations (gliders or particles), including the emergence of {\it spiral guns} that periodically produce mobile localizations.

Some computing capacities and the fundamental complex activity of the Spiral rule are introduced in \cite{kn:AW06}. Besides, a summary of complex structures, collisions, and basic properties of the Spiral rule can be found on Wuensche's home page (published as well in \cite{kn:AWC06}).\footnote{Spiral rule home page: \url{http://www.sussex.ac.uk/Users/andywu/multi_value/spiral_rule.html}}

Previous results have discussed the universality of hexagonal CA. Morita {\it et al.} \cite{kn:MMI99} developed a Fredkin gate in a reversible hexagonal partitioned CA characterised by the conservation of particles and elastic collisions. Adachi {\it et al.} \cite{kn:APL04} implemented delay-insensitive circuits on asynchronous totalistic CA working with additive functions as well. Maydwell \cite{kn:Mayd07} presented interesting variations of this kind of systems known as hexagonal Life-like rules. 

Therefore, the study of hexagonal CA is a work in progress. The contribution of this paper is the implementation of universal logic gates in the Spiral rule through collisions among mobile localizations produced by spiral guns. In addition, an improved classification of complex structures is given, reporting new complex patterns.

The paper is organised as follow: Section 2 describes the Spiral rule automaton. Section 3 presents an extended analysis of the complex patterns in the Spiral rule. Finally, Sec. 4 discusses the implementation of collision-based logic gates.

\section{The Spiral rule CA}

\begin{figure}[th]
\centering
\subfigure[]{\scalebox{0.18}{\includegraphics{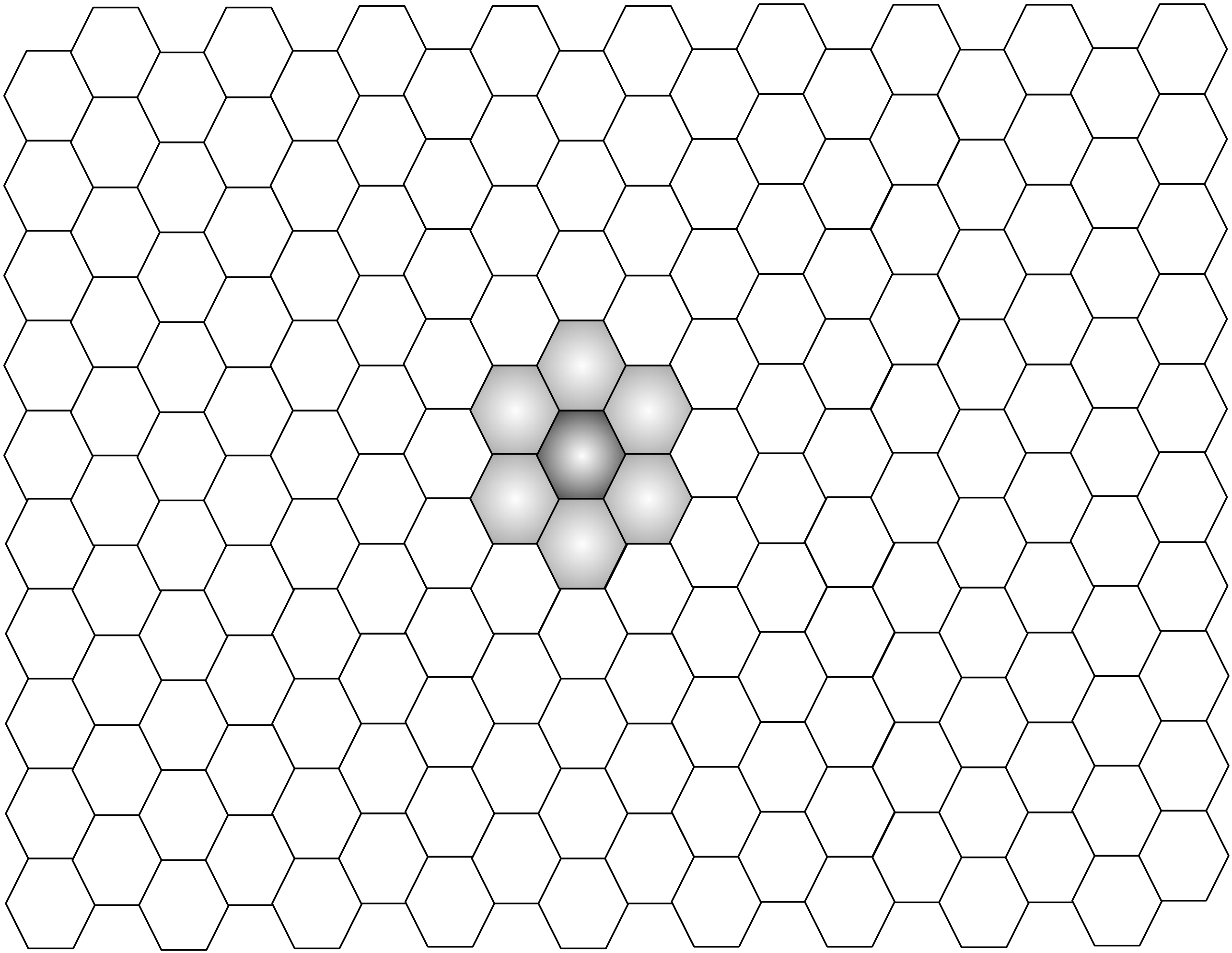}}} \hspace{1.8cm}
\subfigure[]{\scalebox{0.45}{\includegraphics{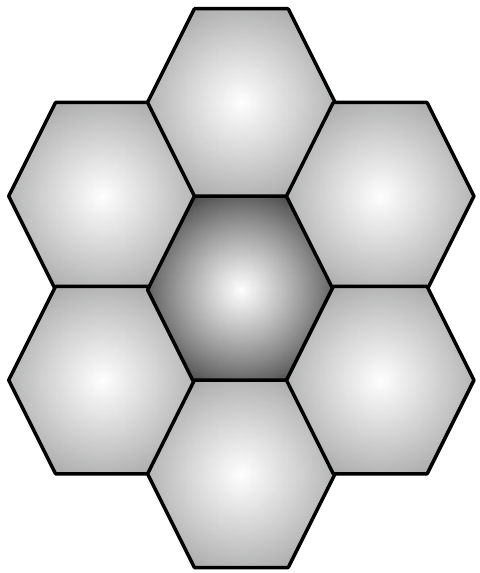}}}
\caption{Hexagonal lattice and a 7-neighbourhood respectively.}
\label{lattice}
\end{figure}

The Spiral rule is a two-dimensional, three-state CA evolving on a hexagonal lattice (Fig.~\ref{lattice}a). The hexagonal local function $f$ is a variation of Moore's function as an isotropic $\cal V=$7-neighbourhood \cite{kn:Wain71} (Fig.~\ref{lattice}b):

\begin{equation}
{\footnotesize
f({\cal V})^t \rightarrow x_{i,j}^{t+1}
}
\end{equation}

Let $\Sigma = \{0,1,2\}$ be the alphabet, so the local function $f : \Sigma^{\cal V} \rightarrow \Sigma$ takes into account $k^{\cal V}=2187$ neighbourhoods, where $k=|\Sigma|$. However, the Spiral rule is a totalistic CA which compresses the number of neighbourhoods by the sum of their states \cite{kn:Wolf83}. Thus the Spiral rule is coded by the number of cells in $\Sigma_i$ on $\cal V$ as shows Table~\ref{totalistic_SR}.

\begin{table}[th]
\centering
\caption{Totalistic function for Spiral rule}
\footnotesize
\begin{tabular}{cc}
766555444433333222222111111100000000 & $\Sigma_2$ \\
010210321043210543210654321076543210 & $\Sigma_1$ \\
001012012301234012345012345601234567 & $\Sigma_0$ \\
\cline{1-1}
000200120021220221200222122022221210 & 
\end{tabular}
\label{totalistic_SR}
\end{table}

This means; for example, that given a neighbourhood with seven cells in state 2 and none in state 1 or 0, it evolves into 0 in the next generation. For simplicity, we can represent the totalistic code in hexadecimal notation as 020609a2982a68aa64; although looking for a more transparent representation, the totalistic evolution rule can be represented as a triangular matrix (Table~\ref{matrix_SR}).

\begin{table}
\centering
\caption{Matrix representation of Spiral rule function}
\footnotesize
\begin{tabular}[h!]{c c | c c c c c c c c}
 & & & & &$|\Sigma_1|$& & & \\
 &&0&1&2&3&4&5&6&7 \\
\cline{2-10}
 &0&0&1&2&1&2&2&2&2 \\
 &1&0&2&2&1&2&2&2& \\
 &2&0&0&2&1&2&2& & \\
$|\Sigma_2|$&3&0&2&2&1&2& & & \\
 &4&0&0&2&1& & & & \\
 &5&0&0&2& & & & & \\
 &6&0&0& & & & & & \\
 &7&0& & & & & & & \\
\end{tabular}
\label{matrix_SR}
\end{table}

This matrix describes the number of cells in state 1 as columns, the number of cells in state 2 as rows and, the number of cells in state 0 is deduced by $7-(|\Sigma_1|+|\Sigma_2|)$. For example, whether we have three cells in state 2 and two cells in state 1, there are two cells in state 0 and the neighbourhood evolves into state 2.

\section{Complex dynamics in the Spiral rule}

The Spiral rule brings a new universe of complex patterns raising in a hexagonal evolution space. In this section, it is presented a number of new structures on the Spiral rule CA. Eventually, such complex patterns become very useful to develop computing devices, or potentially for other engineering devices indeed.

\subsection{Mobile localizations: {\it gliders}}
The Spiral rule has a great diversity of gliders travelling in the evolution space. These mobile localizations (known as gliders in CA literature), can be described by a number of particular properties as: mass, volume, period, translation, and speed.

\begin{figure}
\centering
\includegraphics [width=\textwidth]{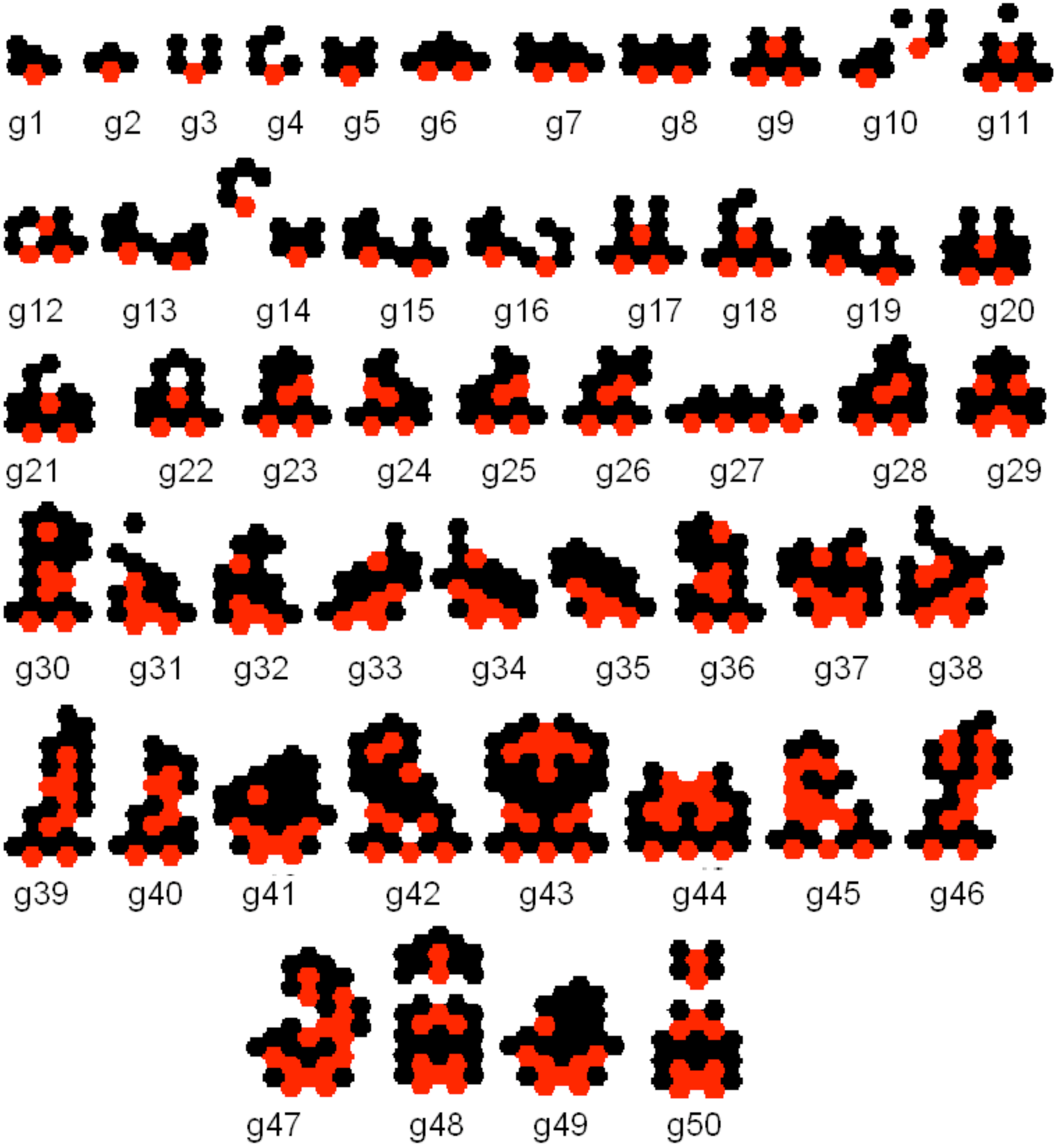}
\caption{Gliders in Spiral rule; state 2 is represented in black, state 1 in red, and the stable background (state 0) in white.}
\label{gliders}
\end{figure}

Up to now, we have enumerated 50 gliders, several of them are new with regard to previous results. Figure~\ref{gliders} displays all the known gliders in the Spiral rule starting from basic or primitive gliders up to large and composed ones (including those which can be extended). Experimentally we have observed that several of them do not have a high probability to emerge from random initial conditions and only survive for few generations, because they are very sensitive to small perturbations.

Table~\ref{gliderTable} depicts general properties for every glider in the Spiral rule; where {\it mass} represents the number of cells in state $1$ and $2$ inside each glider volume. If it has more than one form during its period, the mass is the biggest number of cells. {\it Period} is the number of evolutions needed for each glider to return to the same shape, {\it translation} is the movement measured in cells of a glider during its period; and finally, {\it speed} is calculated as the rate of a glider translation between its period.

\begin{table}[th]
\centering
\caption{Glider properties in Spiral rule. State substrate (state 2), activator (state 1), and inhibitor (state 0).}
\footnotesize
\begin{tabular}[b!]{c|c|c|c|c||c|c|c|c|c}
\hline
glider & mass & period & translation & speed & glider & mass & period & translation & speed \\
\hline \hline
$g_1$ & 5 & 1 & 1 & 1 & $g_{26}$ & 16 & 2 & 2 & 1 \\
\hline
$g_2$ & 4 & 2 & 2 & 1 & $g_{27}$ & 17 & 2 & 2 & 1 \\
\hline
$g_3$ & 5 & 2 & 2 & 1 & $g_{28}$ & 17 & 4 & 4 & 1 \\
\hline
$g_4$ & 5 & 2 & 2 & 1 & $g_{29}$ & 17 & 4 & 4 & 1 \\
\hline
$g_5$ & 6 & 1 & 1 & 1 & $g_{30}$ & 18 & 4 & 4 & 1 \\
\hline
$g_6$ & 8 & 1 & 1 & 1 & $g_{31}$ & 18 & 4 & 4 & 1 \\
\hline
$g_7$ & 9 & 1 & 1 & 1 & $g_{32}$ & 19 & 4 & 4 & 1 \\
\hline
$g_8$ & 10 & 1 & 1 & 1 & $g_{33}$ & 19 & 8 & 8 & 1 \\
\hline
$g_{9}$ & 10 & 1 & 1 & 1 & $g_{34}$ & 20 & 8 & 8 & 1 \\
\hline
$g_{10}$ & 10 & 4 & 4 & 1 & $g_{35}$ & 22 & 4 & 4 & 1 \\
\hline
$g_{11}$ & 11 & 1 & 1 & 1 & $g_{36}$ & 23 & 4 & 4 & 1 \\
\hline
$g_{12}$ & 11 & 4 & 4 & 1 & $g_{37}$ & 24 & 4 & 4 & 1 \\
\hline
$g_{13}$ & 11 & 4 & 4 & 1 & $g_{38}$ & 25 & 4 & 4 & 1 \\
\hline
$g_{14}$ & 11 & 4 & 4 & 1 & $g_{39}$ & 25 & 8 & 8 & 1 \\
\hline
$g_{15}$ & 11 & 4 & 4 & 1 & $g_{40}$ & 26 & 8 & 8 & 1 \\
\hline
$g_{16}$ & 11 & 4 & 4 & 1 & $g_{41}$ & 29 & 4 & 4 & 1 \\
\hline
$g_{17}$ & 12 & 1 & 1 & 1 & $g_{42}$ & 29 & 8 & 8 & 1 \\
\hline
$g_{18}$ & 12 & 2 & 2 & 1 & $g_{43}$ & 31 & 4 & 4 & 1 \\
\hline
$g_{19}$ & 12 & 4 & 4 & 1 & $g_{44}$ & 31 & 4 & 8 & 1 \\
\hline
$g_{20}$ & 14 & 2 & 2 & 1 & $g_{45}$ & 32 & 4 & 4 & 1 \\
\hline
$g_{21}$ & 14 & 2 & 2 & 1 & $g_{46}$ & 32 & 4 & 4 & 1 \\
\hline
$g_{22}$ & 14 & 2 & 2 & 1 & $g_{47}$ & 36 & 4 & 4 & 1 \\
\hline
$g_{23}$ & 15 & 2 & 2 & 1 & $g_{48}$ & 36 & 4 & 4 & 1 \\
\hline
$g_{24}$ & 16 & 2 & 2 & 1 & $g_{49}$ & 43 & 4 & 4 & 1 \\
\hline 
$g_{25}$ & 16 & 2 & 2 & 1 & $g_{50}$ & 47 & 4 & 4 & 1
\end{tabular}
\label{gliderTable}
\end{table}

With such a diversity of gliders, we can refine a classification taking species of them. Table~\ref{speciesGliders} presents three main branches of species in the Spiral rule. In particular, {\it extended} gliders can be configured and connected as mobile {\it polymers} over their six possible directions.

\begin{table}[th]
\centering
\caption{Species of gliders in the Spiral rule.}
\small
\begin{tabular}[b!]{c|l}
\hline
specie & glider \\
\hline \hline
primitive & $g_1$, $g_2$, $g_3$, $g_4$, $g_5$, $g_{29}$ \\
\hline
compound & $g_6$, $g_7$, $g_9$, $g_{10}$, $g_{12}$, $g_{13}$, $g_{14}$, $g_{15}$, $g_{16}$, $g_{17}$, $g_{19}$, $g_{27}$, $g_{35}$\\
\hline
& $g_8$, $g_{11}$, $g_{18}$, $g_{20}$, $g_{21}$, $g_{22}$, $g_{23}$, $g_{24}$, $g_{25}$, $g_{26}$, $g_{28}$, $g_{30}$, \\
extendible & $g_{31}$, $g_{32}$, $g_{33}$, $g_{34}$, $g_{36}$, $g_{37}$, $g_{38}$, $g_{39}$, $g_{40}$, $g_{41}$, $g_{42}$, $g_{43}$, \\
& $g_{44}$, $g_{45}$, $g_{46}$, $g_{47}$, $g_{48}$, $g_{49}$, $g_{50}$
\end{tabular}
\label{speciesGliders}
\end{table}

\begin{figure}
\centering
\subfigure[]{\scalebox{0.5}{\includegraphics{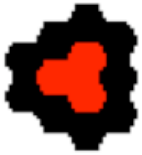}}} \hspace{2.0cm}
\subfigure[]{\scalebox{0.5}{\includegraphics{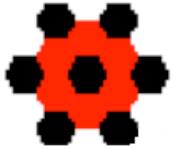}}}
\caption{Still-life configurations in the Spiral rule}
\label{stillLife}
\end{figure}

\begin{figure}
\centering
\subfigure[]{\scalebox{0.43}{\includegraphics{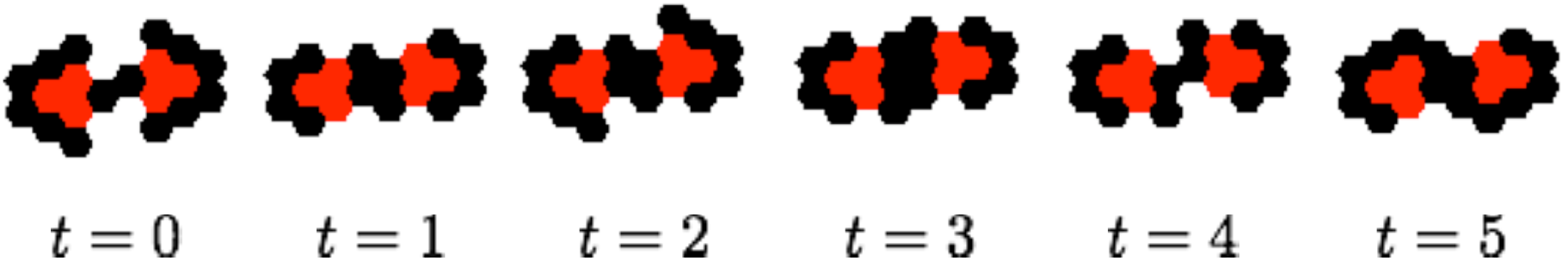}}}
\subfigure[]{\scalebox{0.43}{\includegraphics{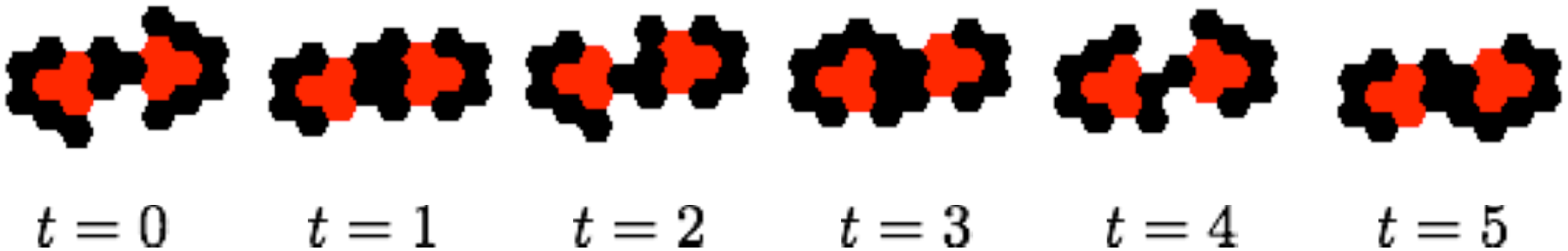}}}
\subfigure[]{\scalebox{0.43}{\includegraphics{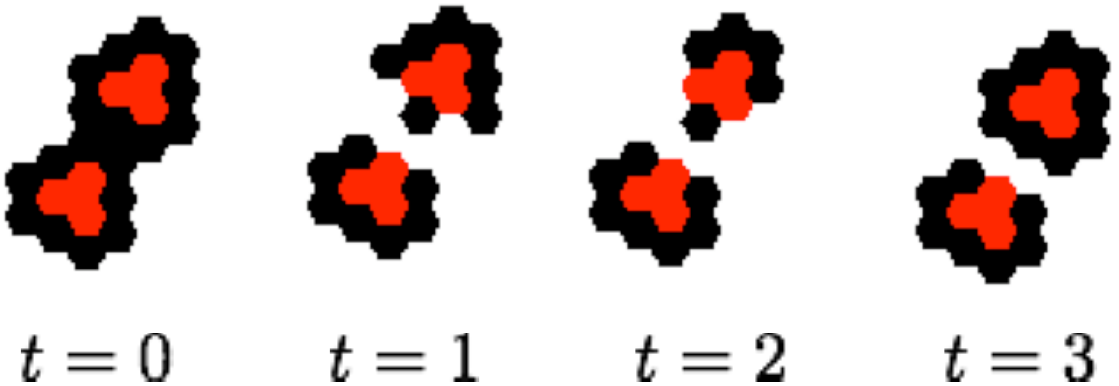}}} \hspace{4.0cm}
\subfigure[]{\scalebox{0.43}{\includegraphics{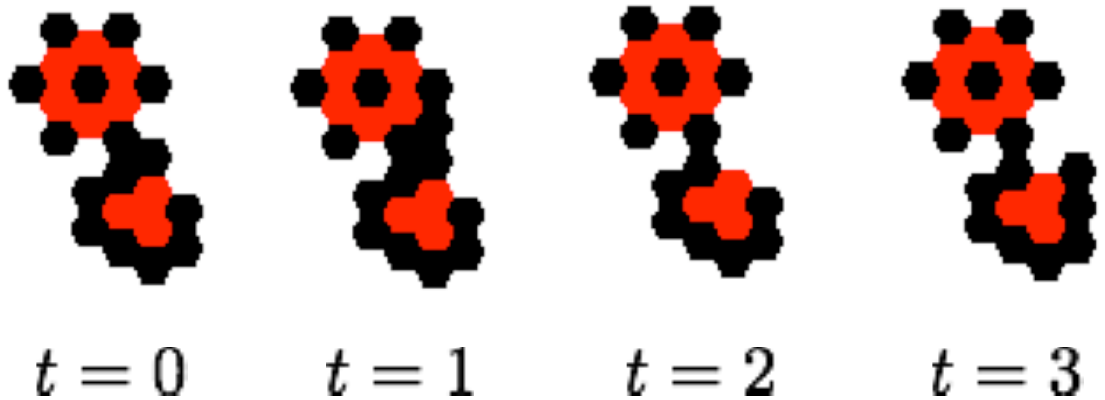}}} \hspace{4.0cm}
\subfigure[]{\scalebox{0.43}{\includegraphics{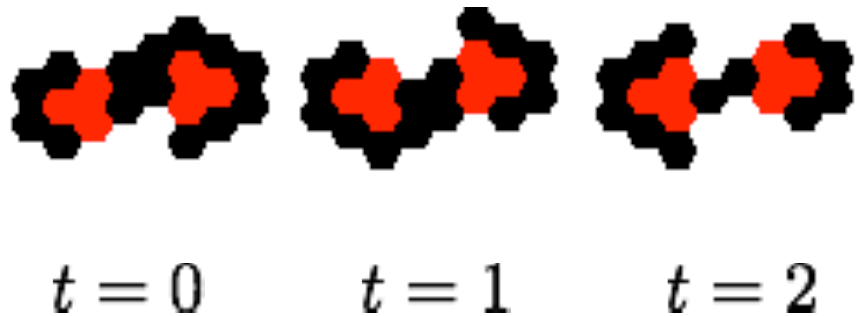}}}
\caption{Oscillating configurations in the Spiral rule.}
\label{oscillators}
\end{figure}

\subsection{Stationary localizations: {\it Still-life configurations}}
Spiral rule has basically a pair of basic or primitives stationary localizations known as still-life configurations in CA. Such patterns can live on the evolution space without alteration; of course, in the lack of any perturbation. Figure~\ref{stillLife} displays these still-life patterns that can be connected as polymers as well to produce extensions of such patterns.

Firstly, the still life `e1' (Fig.~\ref{stillLife}a) has a mass of 12 active cells while the second still life `e2' (Fig.~\ref{stillLife}b) has a mass of 13 active cells. The last still life can be used as a counter of binary strings for a memory device \cite{kn:AMZ10,kn:Zha10}, producing a family of still-life configurations.

A remarkable characteristic (similar to Life) is that both still-life configurations work as ``eaters.'' An eater is a configuration which generally deletes gliders coming from a given direction. This structure eventually becomes very useful to control a number of signals or values in a specific process. For example, deleting values in a computation, where some bits are not needed anymore.

\subsection{Periodic stationary localizations: {\it Oscillators}}
Oscillating patterns are able to emerge in the Spiral rule as well; we can see here an interesting diversity of stationary oscillating patterns. They are frequently a composition of still-life configurations turning {\sc on} and {\sc off} bits periodically.

Figure~\ref{oscillators} presents five kinds of oscillators in the Spiral rule. They are composed by connected still-life configurations, all of them oscillating and changing few values in their structures. Thus, it is not complicated to develop more extended and complex oscillators in the Spiral rule.

\begin{table}[th]
\centering
\caption{Oscillator properties in the Spiral rule: o1 (a), o2 (b), o3 (c), o4 (d), and o5 (e).}
\small
\begin{tabular}[b!]{c|c|c}
\hline
oscillator & mass & period \\
\hline \hline
$o_1$ & 20 & 6 \\
\hline
$o_2$ & 20 & 6 \\
\hline
$o_3$ & 24 & 4 \\
\hline
$o_4$ & 24 & 4 \\
\hline
$o_5$ & 20 & 3
\end{tabular}
\label{oscillatorsProperties}
\end{table}

Table~\ref{oscillatorsProperties} shows general properties for each oscillator in Fig.~\ref{oscillators}. Additionally these oscillators are capable to work as eater configurations as well. However, no simple blinkers or flip-flop configurations are still reported.

\subsection{Glider guns}

\begin{figure}
\centerline{\includegraphics[width=4.4in]{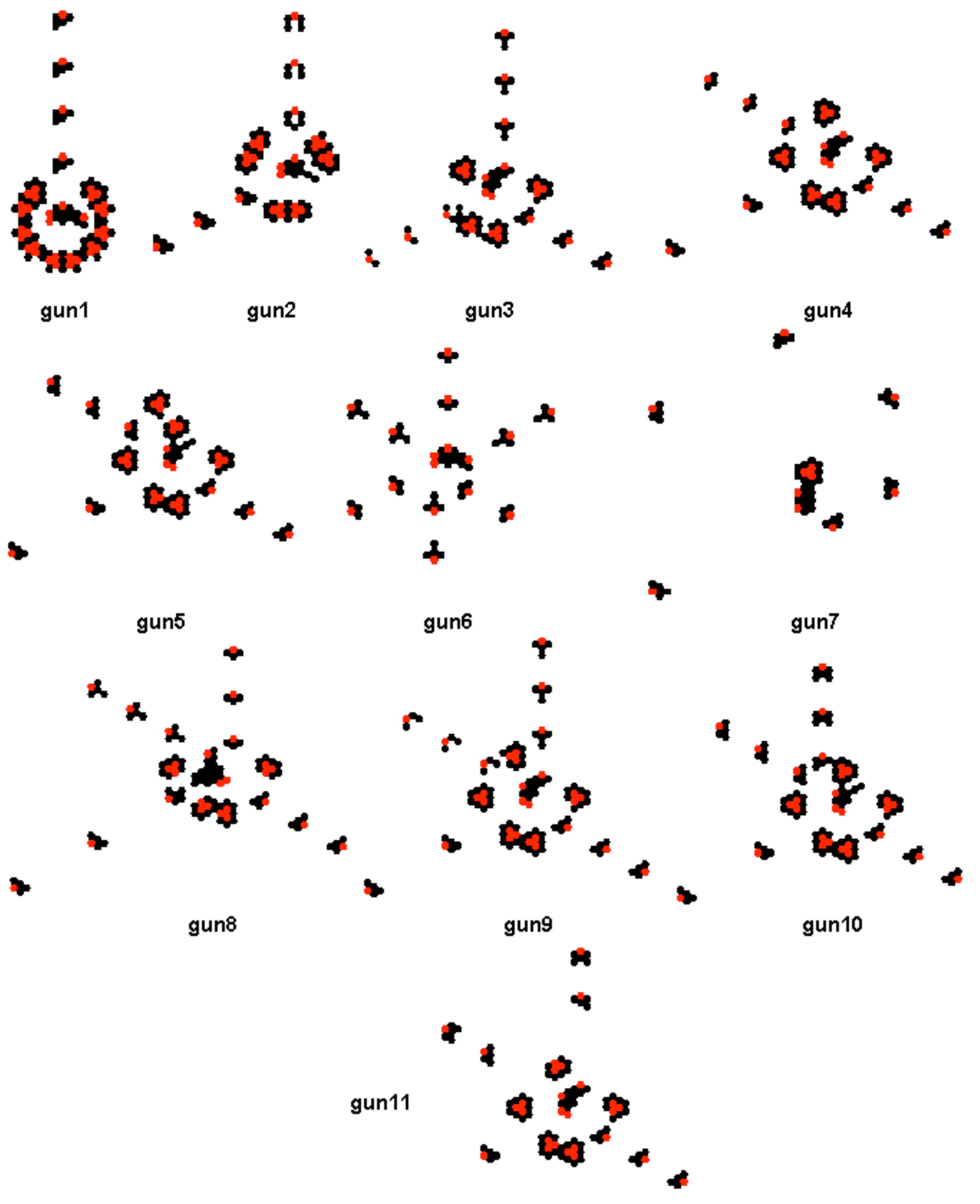}}
\caption{Stationary glider guns in the Spiral rule. A number of non-natural guns are presented too.}
\label{sguns}
\end{figure}

\begin{figure}
\centerline{\includegraphics[width=4in]{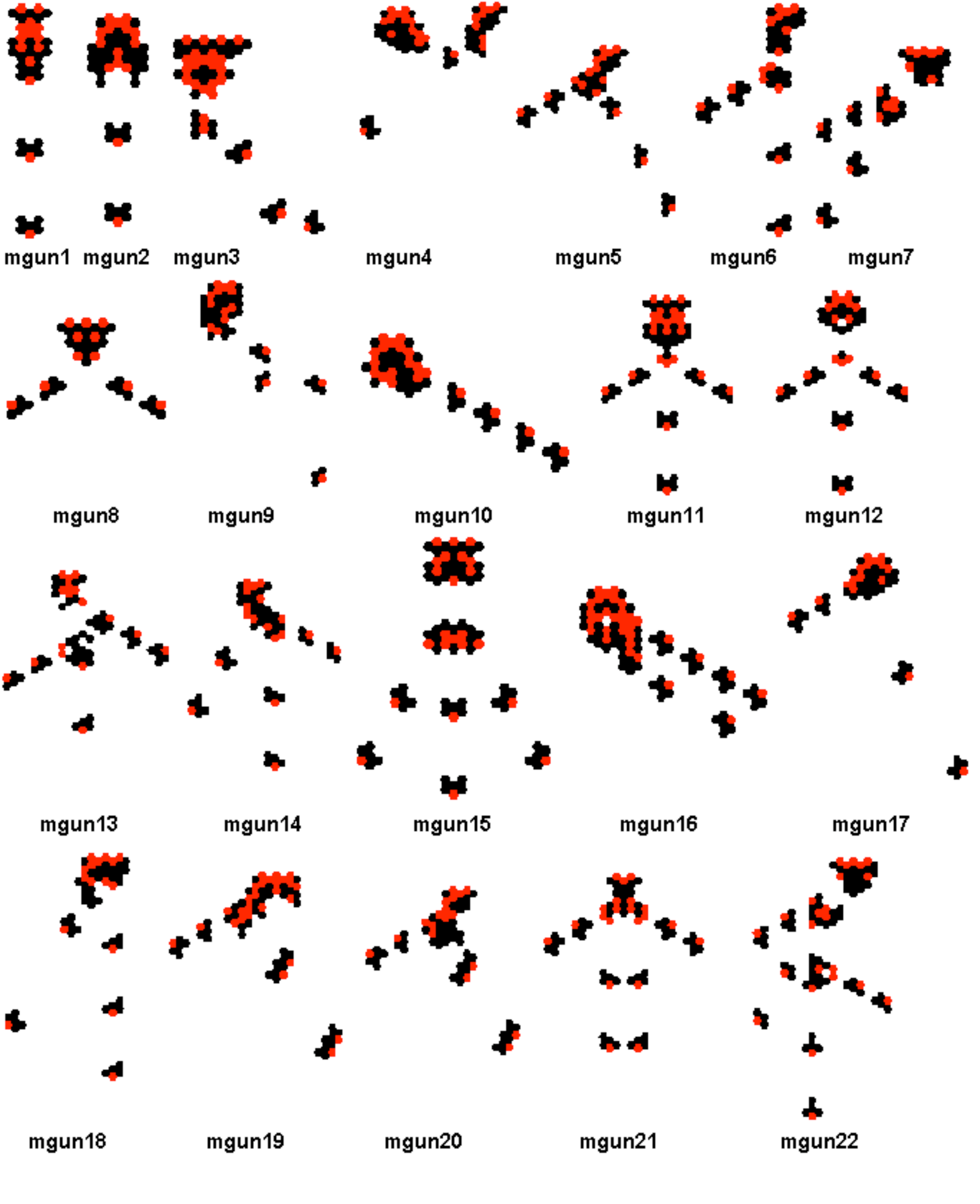}}
\caption{Mobile glider guns in the Spiral rule (first set).}
\label{mgunA}
\end{figure}

\begin{figure}
\centerline{\includegraphics[width=4.4in]{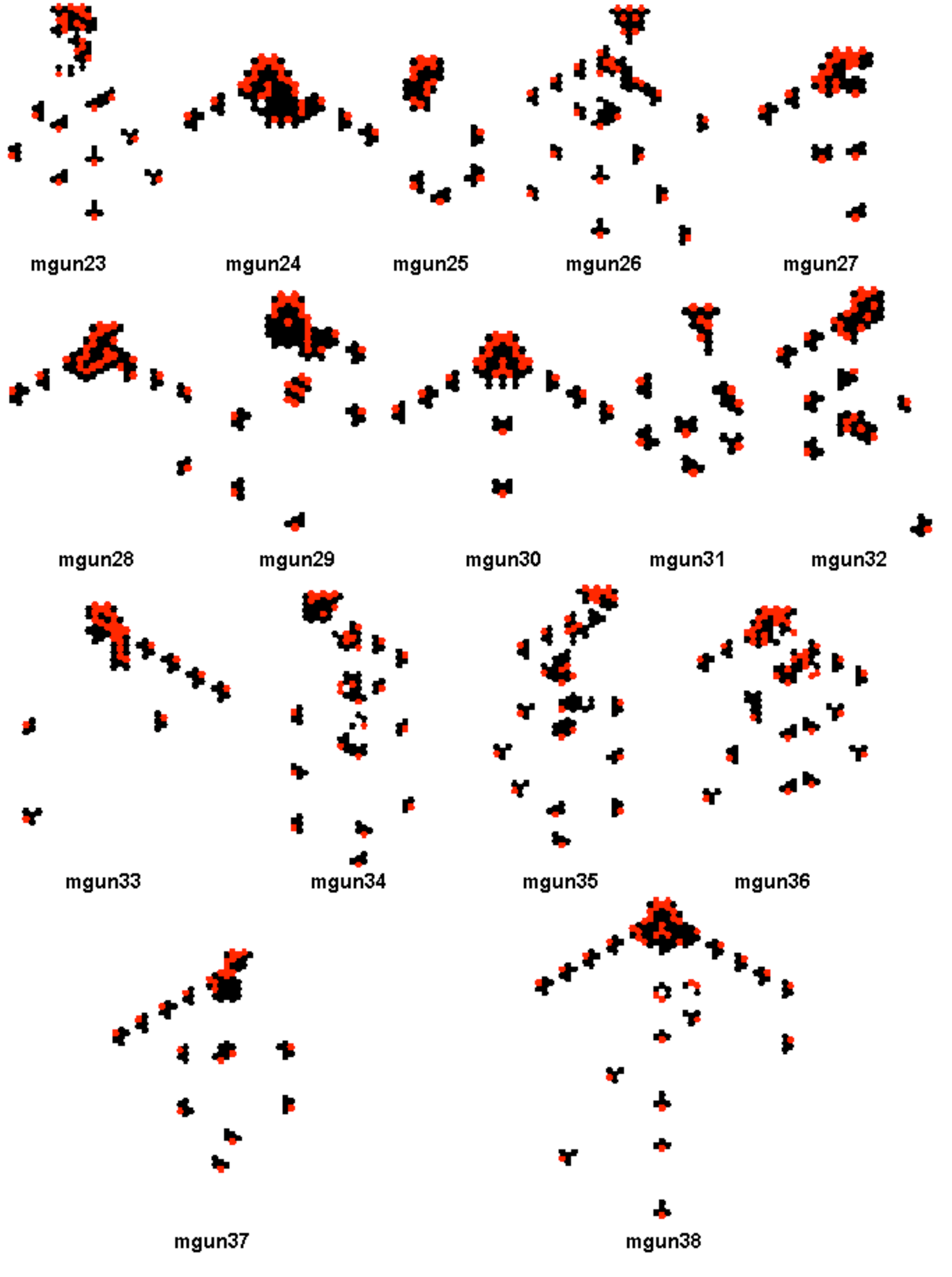}}
\caption{Mobile glider guns in the Spiral rule (second set).}
\label{mgunB}
\end{figure}

One of the most notable features in the Spiral rule CA is the diversity of glider guns that may appear. A {\it glider gun} is a complex configuration generating gliders periodically. In the CA literature, the existence of a glider gun also represents the solution of the unlimited-growth problem \cite{kn:BCG82}.

The Spiral rule has two types of glider guns: stationary and mobile. A stationary gun cannot change of place and position, while a mobile gun can travel along some direction emitting gliders in its path.

Table~\ref{spiralGunsProperties} and Fig.~\ref{sguns} show general properties and dynamics of stationary guns evolving in the Spiral rule respectively. The most frequent glider guns produced by the Spiral rule from random initial conditions are gun6 and gun7. They have a high and slow frequency emitting six $g_2$ and $g_1$ gliders respectively. While gun6 produces six $g_2$ gliders every six generations, gun7 yields six $g_1$ gliders every 22 generations (see Tab.~\ref{spiralGunsProperties}). Other gun variations are obtained adding still life or oscillators, affecting the production of gliders or changing their identity and number. Of course, they are not ``natural'' guns but they can be modified to yield a different number or kind of gliders and its frequency as well, see guns gun1--gun5, gun8--gun11 to look modified guns.

\begin{table}[th]
\centering
\caption{Stationary glider guns properties in the Spiral rule.}
\small
\begin{tabular}[b!]{c|c|c|c|c|c}
\hline
gun & production & frequency & period & volume & gliders \\
& & & & & emitted \\
\hline \hline
gun1 & $g_1$ & 1 & 6 & $15\!\!\times\!\!15$ & 1 \\
\hline
gun2 & $g_1$, $g_3$ & 2 & 6 & $15\!\!\times\!\!15$ & 2 \\
\hline
gun3 & $g_1$, $g_2$, $g_3$ & 3 & 6 & $14\!\!\times\!\!15$ & 3 \\
\hline
gun4 & $3g_1$, $2g_2$ & 5 & 12 & $16\!\!\times\!\!17$ & 3 \\
\hline
gun5 & $5g_1$ & 5 & 12 & $19\!\!\times\!\!17$ & 3 \\
\hline
gun6 & $6g_2$ & 6 & 6 & $8\!\!\times\!\!9$ & 6 \\
\hline
gun7 & $6g_1$ & 6 & 22 & $12\!\!\times\!\!12$ & 6 \\
\hline
gun8 & $3g_1$, $4g_2$ & 7 & 12 & $14\!\!\times\!\!14$ & 4 \\
\hline
gun9 & $3g_1$, $2g_2$, $2g_4$  & 7 & 12 & $15\!\!\times\!\!17$ & 4 \\
\hline
gun10 & $5g_1$, $2g_5$ & 7 & 12 & $15\!\!\times\!\!15$ & 4 \\
\hline
gun11 & $13g_1$, $4g_5$ & 17 & 30 & $15\!\!\times\!\!17$ & 4
\end{tabular}
\label{spiralGunsProperties}
\end{table}

Particularly, stationary glider guns {\it gun6} and {\it gun7} (natural guns in Spiral rule) describe characteristic ``spiral guns'' in chemical reactions, as we can see in Belousov-Zhabotinsky phenomena \cite{kn:ACA05,kn:Ada04,kn:CA05}, gliders are CA analogies to wave-fragments (localised excitations) propagating in sub-excitable reaction.

Also, the Spiral rule has a wide number of mobile glider guns, generally they are formed by complex structures generating more than two gliders. However, the guns are very sensitive to any perturbation, consequently destroying the gun configuration. While natural spiral guns (gun6 and gun7) are very robust to defend their structures from many collisions, there are few of them that are able to destroy these structures as well. Figures~\ref{mgunA} and~\ref{mgunB} present the broad diversity of mobile glider guns in the Spiral rule, having up to 38 different types.

Thus, there is always a way to yield basic gliders in the Spiral rule from some glider gun.

\section{Logic gates and beyond}
This section describes constructions to simulate computing devices in the Spiral rule by glider collisions, implementing universal logic gates and other useful computing devices. They are inspired by previous works in the Game of Life CA \cite{kn:BCG82}. Thus, the presence of gliders represents bits in state 1 and its complement (absence) represents bits in state 0.

\begin{figure}[th]
\centerline{\includegraphics[width=4.3in]{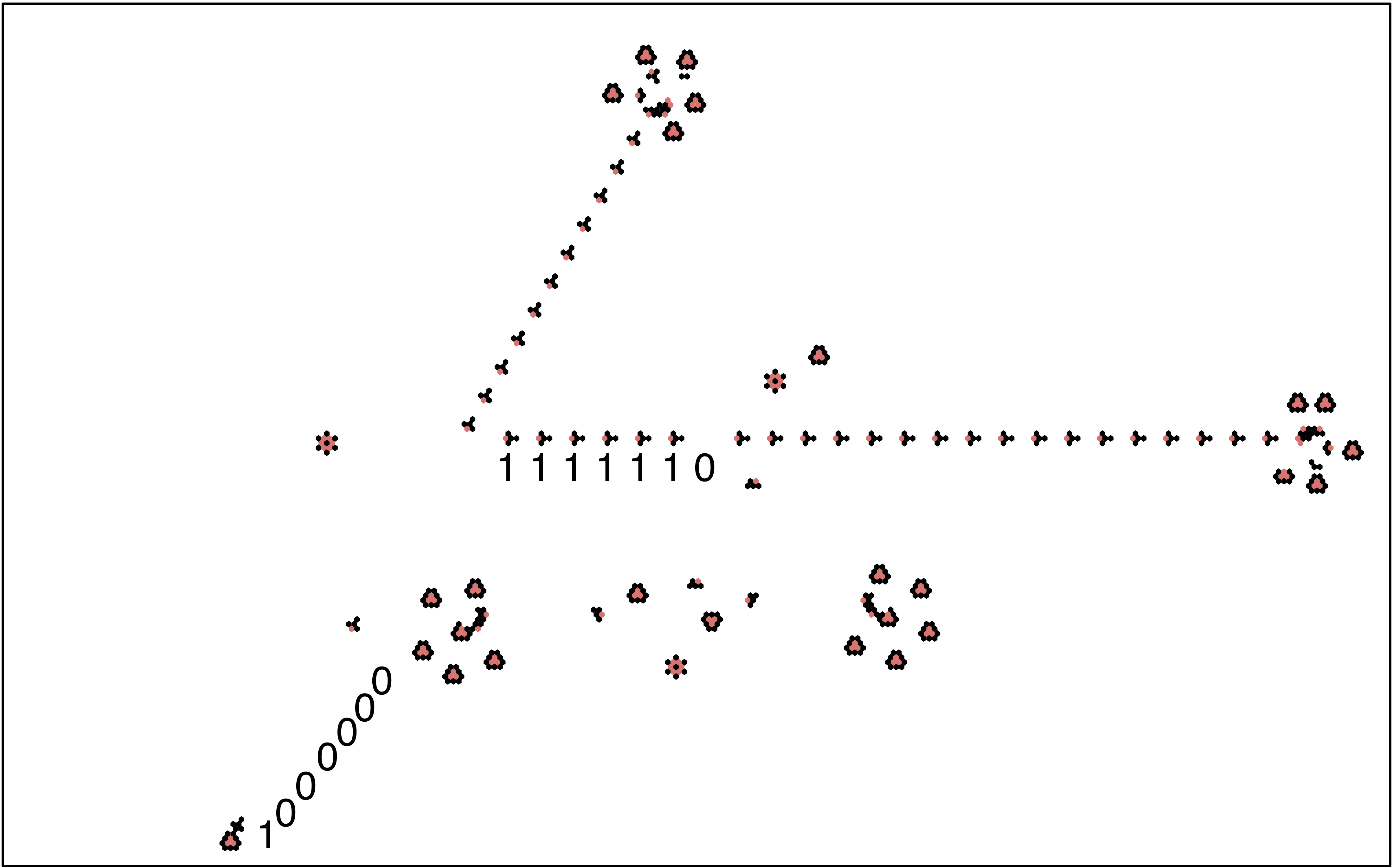}}
\caption{{\sc Not} gate implementation in the Spiral rule.}
\label{notGate-1}
\end{figure}

The first construction implements a {\sc not} gate.\footnote{You can see an animation in \url{http://www.youtube.com/watch?v=_bC5ucq_sKc}. The construction was prepared using DDLab software; the file `notGt\_sr.eed' can be download from \url{http://uncomp.uwe.ac.uk/genaro/Papers/Thesis.html}.} This one presents a {\sc not} gate processing the string $\neg(1111110)$. Here a high frequency spiral gun gun6 produces six gliders where five localizations are suppressed by eaters to preserve only one. Thus, the first spiral gun (east position) yields periodically the sequence $1111\ldots$. Then, other two low frequency spiral guns gun7 generate additional eaters to delete a bit of such sequence, given the string $(1111110)^*$. Finally a fourth spiral gun gun6 (north position) produces the {\sc not} operation obtaining the string $(0000001)^*$ by annihilation reactions. Gaps amongst spiral guns can be manipulated to get a desired string. 

\begin{figure}[th]
\centerline{\includegraphics[width=4.3in]{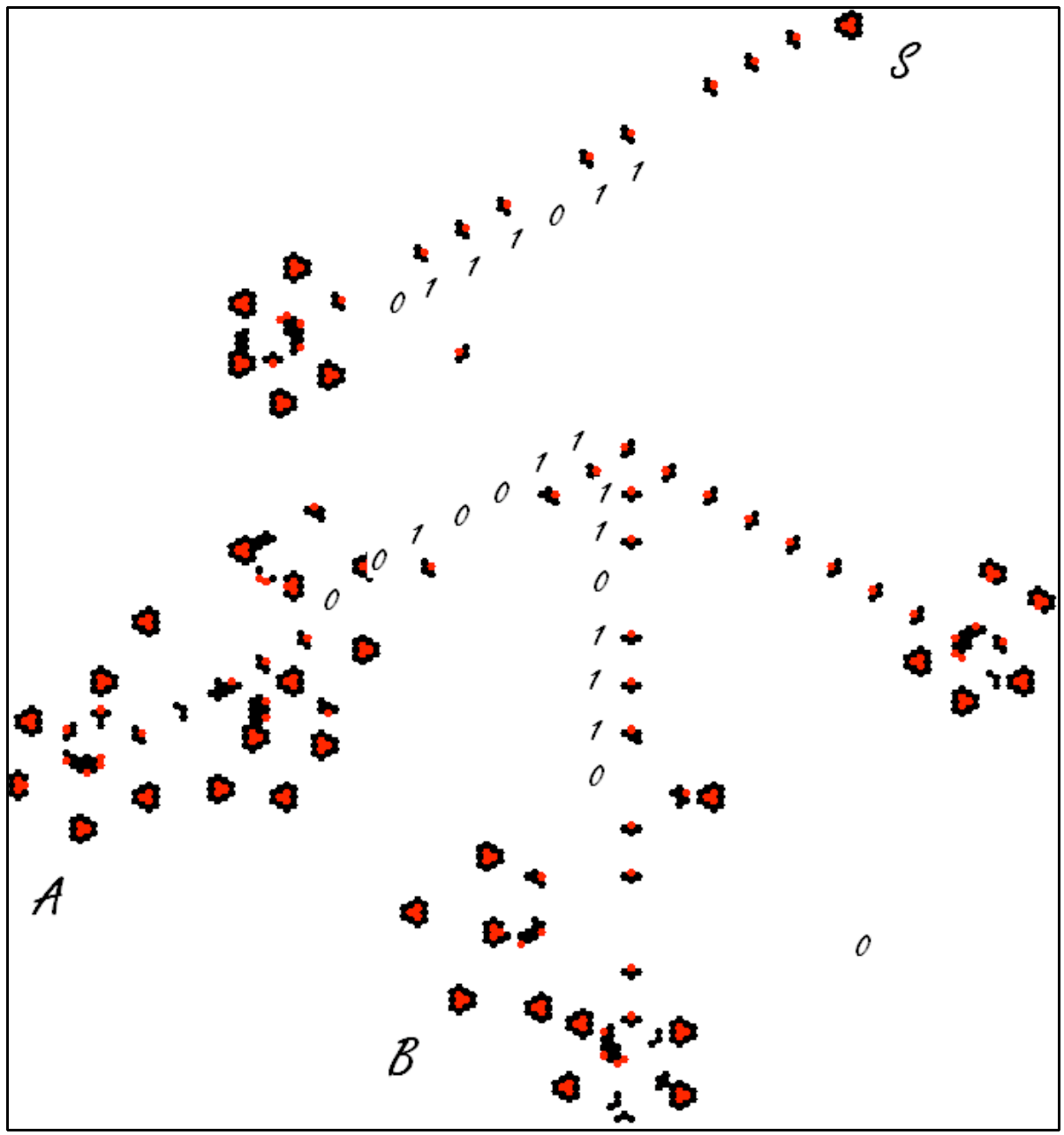}}
\caption{{\sc Or} gate implemented in the Spiral rule. $S=A \mbox{ \sc or } B$.}
\label{orGate}
\end{figure}

\begin{figure}[th]
\centerline{\includegraphics[width=3.2in]{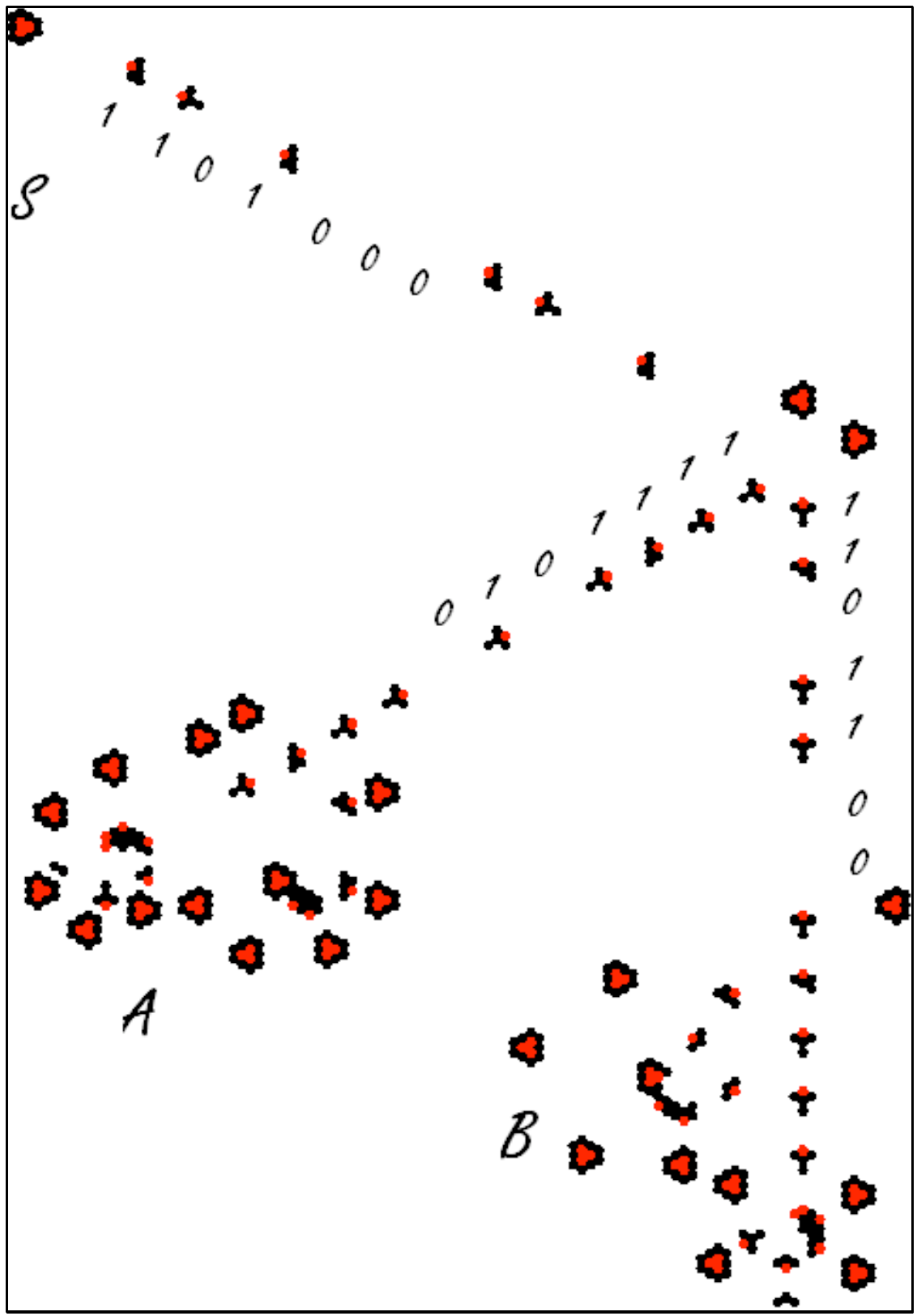}}
\caption{{\sc And} gate implemented in the Spiral rule. $S=A \mbox{ \sc and } B$.}
\label{andGate}
\end{figure}

\begin{figure}[th]
\centerline{\includegraphics[width=4.2in]{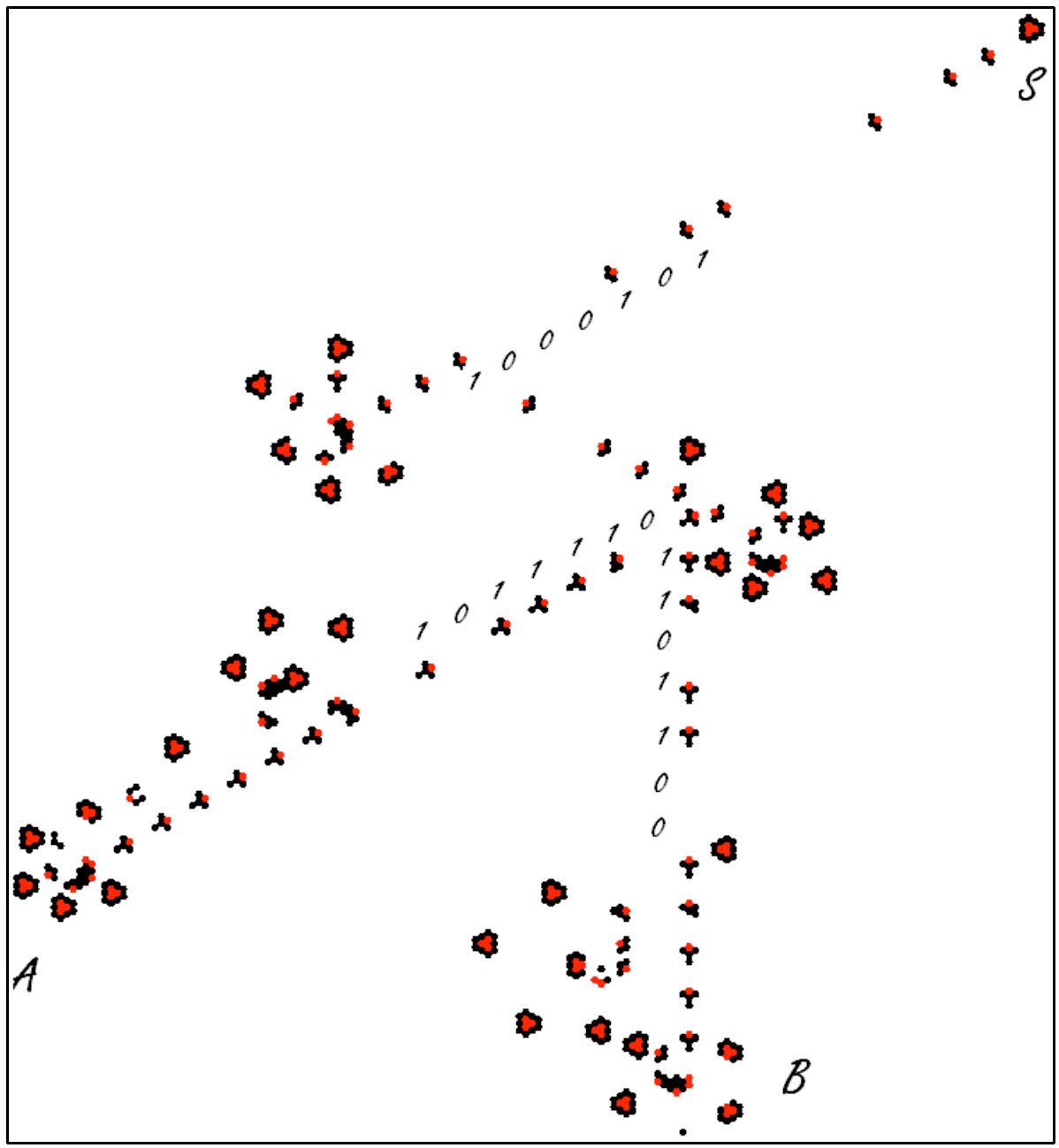}}
\caption{{\sc Xor} gate implemented in the Spiral rule. $S=A \mbox{ \sc xor } B$.}
\label{xorGate}
\end{figure}

\begin{figure}[th]
\centerline{\includegraphics[width=4.2in]{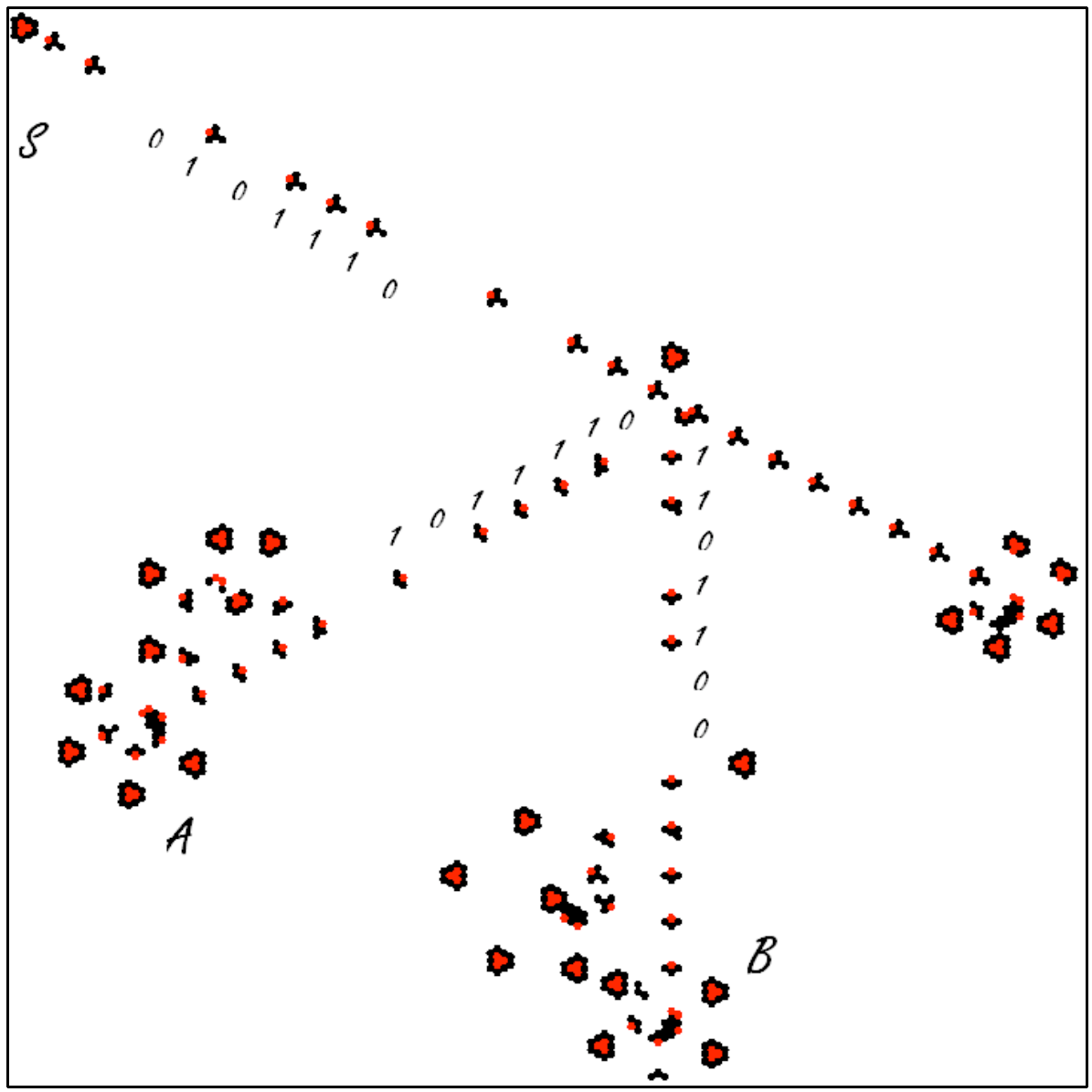}}
\caption{{\sc Xnor} gate implemented in the Spiral rule. $S=A \mbox{ \sc xnor } B$.}
\label{xnorGate}
\end{figure}

In this sense, specific initial conditions have been specified to simulate: {\sc or} (Fig.~\ref{orGate}) and {\sc and} (Fig.~\ref{andGate}) logic gates. Complicated designs are also developed to implement {\sc xor} (Fig.~\ref{xorGate}) and {\sc xnor} (Fig.~\ref{xnorGate}) gates respectively. In these cases, additional still life patterns and spiral guns are needed to synchronise multiple collisions, and controlling sequences of bits.

\begin{figure}
\centering
\subfigure[]{\scalebox{0.4}{\includegraphics{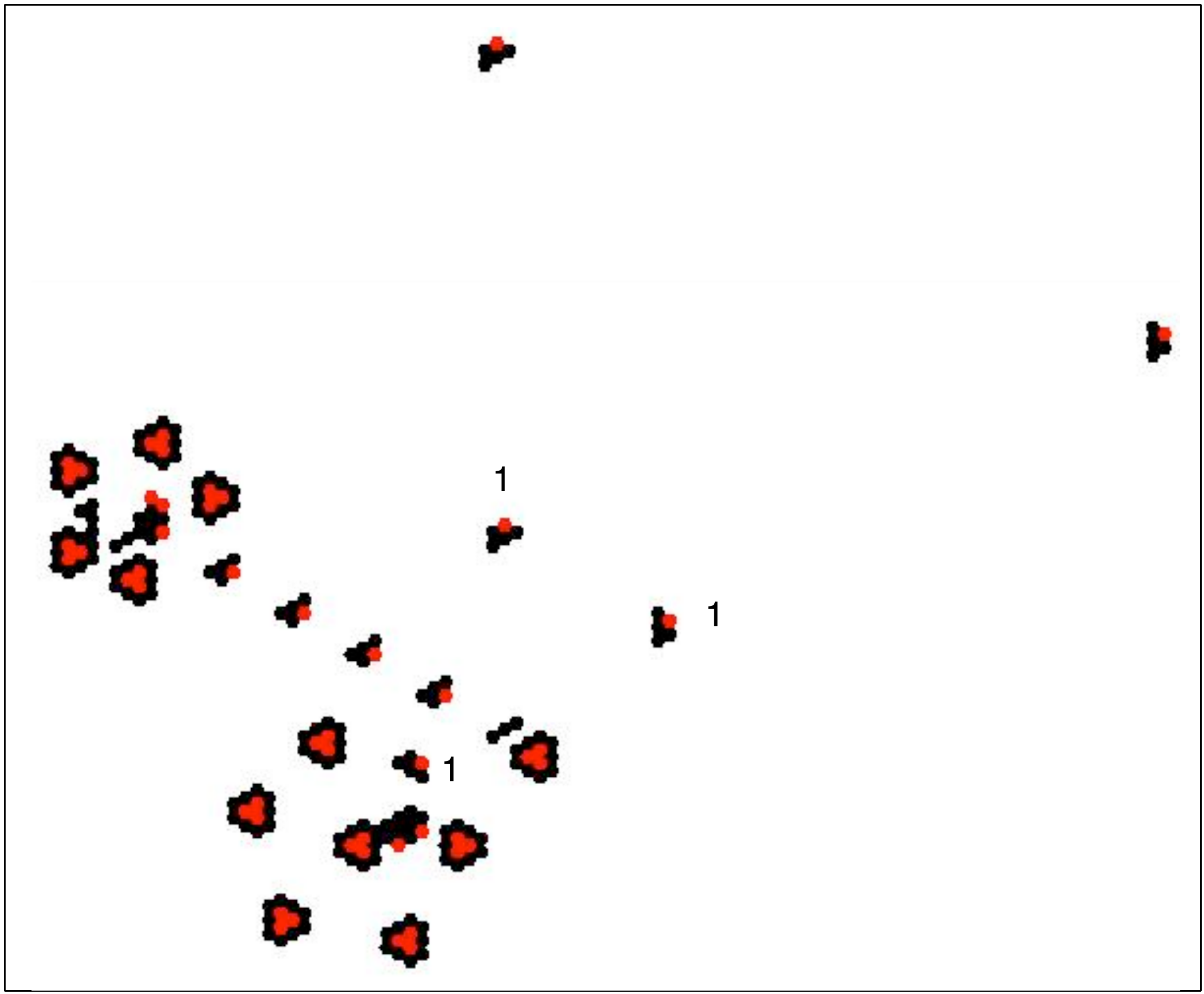}}} 
\subfigure[]{\scalebox{0.4}{\includegraphics{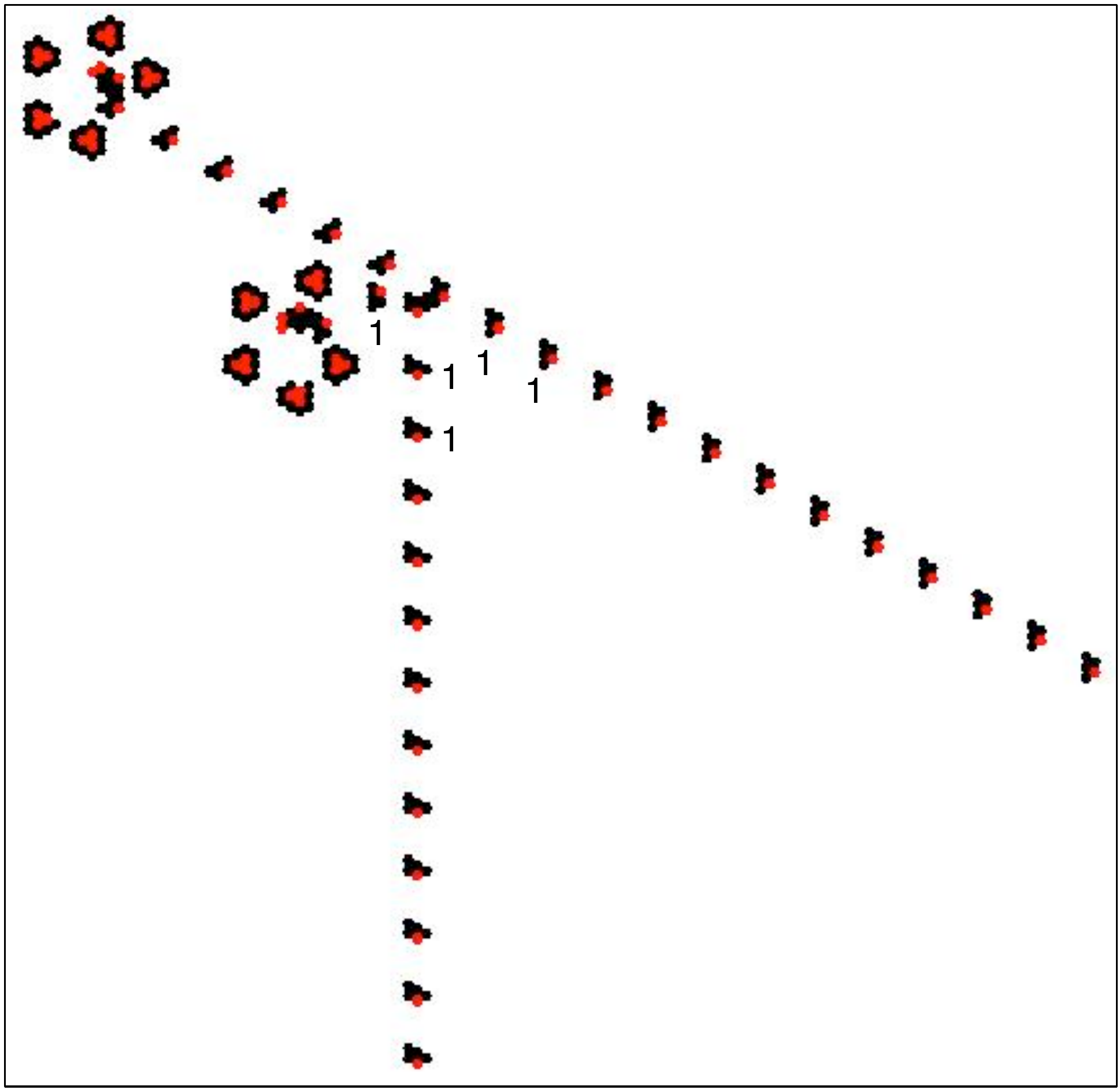}}}
\caption{{\sc Fanout} gates implemented in the Spiral rule.}
\label{fanoutGates}
\end{figure}

Additionally, Fig.~\ref{fanoutGates} displays two {\sc fanout} gates. The first {\sc fanout} gate (Fig.~\ref{fanoutGates}a) is implemented using a new high frequency spiral gun, the gun1 (producing a glider every six times) modified to generate $g_1$ gliders. Then, another gun1 is put at 90 degrees to obtain a multiplication of the original localisation, splitting the first $g_1$ glider in two $g_1$ gliders. The second design depicts a high frequency {\sc fanout} gate (Fig.~\ref{fanoutGates}b) inspired from the spiral gun gun1.

\begin{figure}[th]
\centerline{\includegraphics[width=4.2in]{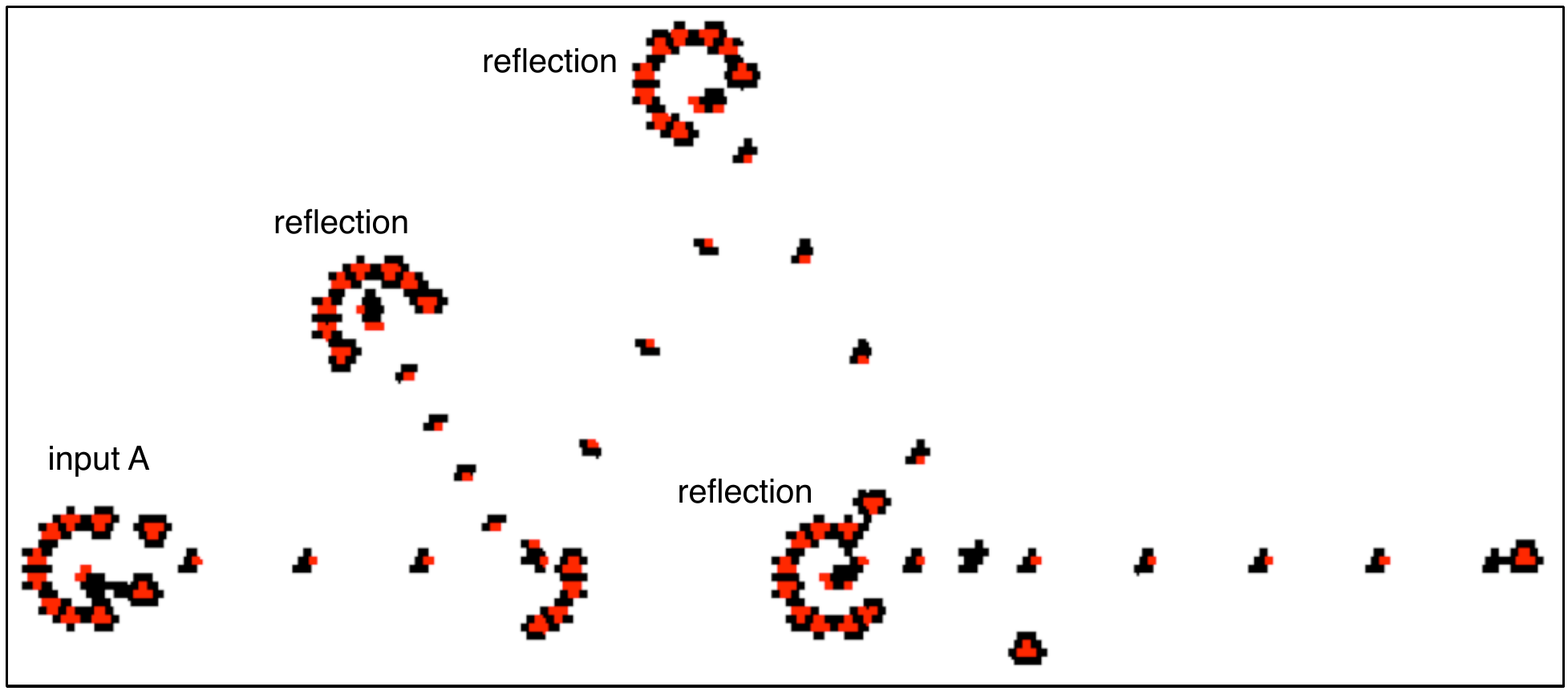}}
\caption{{\sc Delay} device implemented in the Spiral rule.}
\label{delay}
\end{figure}

A low frequency {\sc delay} device (see Fig.~\ref{delay}) has been designed as well, using a spiral gun gun1. The goal is to reflect the original input by three reflection reactions which preserve the same sequence. The delay time can be manipulated increasing the gap between each reflection. Such device could be useful to synchronise multiple signals and generate sophisticated computations in further works.

\section{Final remarks}
Universal logic gates and other computing devices in the Spiral rule have been implemented, showing the potential of this hexagonal CA  to organise complex patterns and synchronise multiple collisions.\footnote{Implementations of {\sf capacitor} and {\sf reflection} devices in the Spiral rule can be watched in \url{http://www.youtube.com/watch?v=Cx5QYxvfF9g}, \url{http://www.youtube.com/watch?v=Cx5QYxvfF9g}, and \url{http://www.youtube.com/watch?v=H2xvG-UHM9o}.} The next step will be the design of full logic circuits working together to get a complete computable function. To reach such constructions, we will develop a more extended engineering based-collision of gliders in the Spiral rule, to get a full universality employing similar constructions specified in other hexagonal CA models \cite{kn:MMI99,kn:APL04}.

About unconventional computing, these results may be useful as a guide to implement reaction-diffusion computers on Belousov-Zhabotinsky systems \cite{kn:ACA05}. Here, mobile localizations are represented as a fragment of waves and their interactions are a scheme for three states: substrate (state 2), activator (state 1), and inhibitor (state 0). The spiral guns represent a discrete version of a classical spiral wave in an excitable medium. An interesting study determining spiral forms in CA can be consulted in \cite{kn:Gord66}. The spiral guns can also be related to crystallisation computers \cite{kn:Ada09} where a crystallised way will be precisely a glider travelling on such direction. Experimental laboratory tests are working in this direction at the ICUC.\footnote{International Centre of Unconventional Computing, University of the West of England, Bristol, United Kingdom. Home page \url{http://uncomp.uwe.ac.uk/}.}

All simulations were done with {\sf SpiralSimulator}\footnote{Here you can download {\sf SpiralSimulator} software and source files to reproduce every logic gate designed in this paper \url{http://uncomp.uwe.ac.uk/genaro/Papers/Thesis.html}.}, and {\sf DDLab}\footnote{Here you can download {\sf DDLab} software \url{http://www.ddlab.org/}.} free software \cite{kn:Wue10}.

\section*{Acknowledgement}
We thanks useful discussions to Wuensche and Adamatzky that help us to improve this paper. Rogelio B. and Paulina A. L. thanks to support given for ESCOM-IPN, CINVESTAV-IPN and CONACYT. Genaro J. M. thanks to support given by EPSRC grant EP/F054343/1. Juan C. S.-T.-M. thanks to support given by CONACYT through project number CB-2007-83554.


\end{document}